\documentclass[11pt]{revtex4}
\usepackage{amsmath, amssymb, lipsum}
\usepackage{graphics, graphicx, epsfig, color, subfigure}
\begin{document}

\title{\Large Early inflation induced gravity waves can restrict Astro-Particle physics}

\date{\today}

\author{Shmuel Nussinov}
\affiliation{Raymond and Beverly Sackler School of Physics and Astronomy, 
Tel Aviv University, Tel Aviv, Israel.}
\begin{abstract}
In this paper, we discuss limits on various astro-particle scenarios if the scale \textit{and} the reheat temperature of the last relevant inflation were very high. While the observed ``B" like pattern of polarizations of the CMB suggest a very high ($\ge 10^{16}\ GeV$) scale of a primordial (which motivated this work initially) and may reflect effects of dust, we believe that addressing these issues is nonetheless very useful. We recall the potential difficulties with various topological defects - monopoles, strings and domain walls generated at the SSB (spontaneous symmetry breaking) of various gauge symmetries. The main part of the paper is devoted to discussing difficulties with long-lived heavy particles, which could be dark matter but cannot efficiently annihilate to the required residual density because of basic S-Matrix unitarity/analyticity limits. We indicate in simple terms yet in some detail how the WIMP miracle occurs at $M(X)\sim{TeV}$ and how the axiomatic upper bound presently updated to $M(X) \le{110 TeV}$ was originally derived by Greist and Kamionokowski.  We also argue that generically we expect the stronger $M(X)\le{20\ GeV}$ bound to hold. We then elaborate on the pure particle physics approaches aiming to enhance the annihilation and evade the bounds. We find that the only and in fact very satisfactory way of doing this requires endowing the particles with gauge interactions with a confinement scale lower than $M(X)$. We also comment on models with light $O(KeV)$ dark matter, which was supposed to be frozen in via out-of equilibrium processes so as to have the right relic densities pointing out that in many such cases \textit{very} low reheat temperatures are indeed required and speculate on the large desert scenario of particle physics. Most of what we discuss is not new but was not presented in a coherent fashion.
\end{abstract}

\maketitle

\section{Introduction}
The early universe serves as a unique, hot laboratory for particle physics~\cite{Dolgov, Kolb}. The correctly predicted abundance of light elements suggested three light neutrino flavors before the direct confirmation by $Z$ width measurements. Structure evolution and CMB (cosmic microwave background) observations further limited the sum of the neutrino masses to be smaller than $\sim{eV}$ and the precise Planckian spectrum of the CMB excludes late decay/annihilation of heavy particles which may distort the spectrum.

A most relevant question is ``How hot a laboratory was it" or phrased in the context of Inflation ``What are the energy scale $\Lambda$ and the reheat temperature $T_R$ of the last relevant inflation". A relevant inflation is operationally defined as one with many e- fold expansion sufficient to significantly dilute the number densities of particles or topological structures inherited from previous epochs.

Recent observation by the  BICEP collaboration $\it{if}$ not due to dust, suggest ``B-like" patterns in the CMB polarization.~\cite{Bicep}. To generate it one needs gravity waves emitted early on and with sufficient intensity to be able to imprint at recombination this polarization pattern. The ratio r of tensor to scalar fluctuations $r(Bicep)=0.2$ suggested a high mass scale for the primordial inflation: $ \Lambda(I)=V^{1/4} \approx 2 10^{16} GeV$ , only  two orders of magnitude below the Planck mass and consistent with the scales of  GUT (grand unified theories). Even smaller $r=0.05-0.1$ values along with WMAP and PLANCK measurement of a density fluctuation index $n_s$ close to unity suggest large $\Lambda(I)\ge {10^{15}}GeV$.

The simplest interpretation of these results involves an inflanton of mass $M_{inlaton} \sim{\Lambda(I)}$ with initial Planckian offset away from zero of the inflaton field driving the inflation. We choose  a high inflanton mass in order to avoid extreme super-planckian deviations $\Delta(\phi)$ which would be required in order to have $ V(\phi) \sim {m^2\phi^2}\sim {\Lambda^4}$ . While one can have the desired sixty e-fold horizon expansion and $n_s$ close to unity in many scenarios this would be among the simpler ones. Unless the decaying inflanton prefers certain sectors of particle theory this suggests that all forms of matter have been early on in thermal equilibrium at a reheat temperature $T(R)$ which is not \textit{much} smaller than m(Inflaton) or $\Lambda(I) $. A universal couplings are natural if the inflaton is an alias for quantum gravity~\cite{Brout}.

We can choose a very small $m(Inflaton)/{M (Planck)}$ - or small fine-tuned quartic dimensionless couplings $\lambda$. While we will then need very large, superplanckian $\Delta(\phi)$  to achieve the energy density required, the reheat temperature can be much smaller than $\Lambda$. Still in the following we would like to discuss what would happen if :

(i) the scale $\Lambda$ of the last relevant inflation is high? or

(ii) The reheat temperature $T(R)$ after this inflation is high?

\noindent
The second assumption is much stronger as lower reheat temperatures can be achieved if the coupling of the inflaton to all other forms of matter is very small. While much of the material below is not new, it is useful to emphasize the limitations on many particle physics models which the above assumptions imply.

In Sec II we show how the high inflation scale can constrain models of symmetry breaking leading to topological defects. 

In the main sec III, we recall difficulties of generic CDM (cold dark matter) models with high masses $M(X)$. The bounds originally emphasized by Greist and Kamionokowski (G.K)~\cite{Greist} stem from the small annihilation cross-sections and the resulting excessive residual densities. We review this seminal work focusing on the particle physics entering the cross-section bounds. We argue that the bounds on annihilation cross-sections  are in general stronger and for a more restrictive upper bound than the original $M(X)\le 340 TeV$ of GK. 

In Sec IV we discuss how these bounds dramatically relax if the X's participate in a confining non-abelian gauge interaction with a scale $\Lambda'$ much smaller than $M(X)$ and trace this to the fact that S matrix framework cannot address confinement. In sec V we consider models of light $m(x)\sim {KeV}$ very weakly interacting particles which are assumed to have never been in  equilibrium with the rest of matter/ radiation. In such models one can ``freeze-in" a desired relic density of these particles which may be ``warm" dark matter. However in even modest re-heat temperatures far lower than the very high values considered here these particles which do not decay or annihilate are thermally produced with over-closure densities. 

Finally in the concluding section VI, we comment on a possible connection with the ``large desert" particle physics scenario.

Note that the panacea of diluting the density of defects or of heavy relics by either subsequent inflation or by the late decays of some heavy, long lived particles, such as ``Moduli"- massive scalars which couple only with gravitational strength appearing in many string inspired scenarios - may fail. In later inflations the gravity waves emitted in the primordial high scale inflation get diluted. The same dilution operates also on the CMB photons or more generally the ``radiation" which later converts into photons. However at he end of the later inflation we expect some reheat and consequent generation of many more  photons, but \textit{no} more gravitational waves. Likewise the late decays of massive particles will enhance the photons but \textit{not} the gravitons or gravitational waves This imbalance -and ensuing relative weakening of the gravity wave amplitude- may prevent them from imprinting the B pattern on the photons which BICEP may have seen.

In the following, we do not address these fixes of the problems which arise in many scenarios and focus on possible solution within the particle physics proper. 

\section{Consequence of a high scale of the last inflation for models with topological defects.}
Inflation can dilute excessive amount of topological defects. Indeed modern inflation was motivated by an attempt to dilute the density of supper-heavy GUT magnetic monopoles generated at the first stage of GUT breaking in which a $U(1)$ subgroup splits off: $G_{GUT}\rightarrow G'XU(1)$~\cite{Guth}. Production of monopoles even at energies bigger than their masses is dynamically suppressed by $\exp\left ( -1/{\alpha} \right )$ type factors~\cite{Drukier}. The fact that $M(monopole)=M(GUT)/{\alpha}$ implies that even if the monopoles were thermally produced at the highest temperatures possible $T=T(Reheat)\approx{M(GUT)}$ there would still be an  $\exp\left ( -1/{\alpha} \right )$  Boltzmann suppression. However such monopoles may be produced at the phase transition corresponding to the spontaneous breaking of GUTs via the Kibble mechanism~\cite{Kibble} at zeros of the GUT breaking Higgs condensate. These points are the junctures of domains where the Higgs VeV points in different directions of the gauge group. without subsequent inflations the primordial inflation at a scale of $M(GUT) = 10^{16} GeV$ may not achieve the dilution of $n_{Monopoles}/{n{\gamma}}=10^{-25}$ required to ensure a relic density of the monopoles less than that of the allowed dark matter.  Demanding that the monopoles will not dissipate the galactic magnetic field Yields the much stronger Parker bound on the density of magnetic monopoles. The capture of $m-\bar{m}$ pairs into ``Monopolium" states with ensuing annihilation cannot achieve such a small monopole number density. We discuss related issues in Sec IV.\newline Next in a natural hierarchy are the one dimensional cosmic strings  arising in spontaneosly broken $U(1)$ symmetries. We expect that approximately one closed string per horizon of length $ct=l(horizon)=l(h) $ forms at the time of the SSB (spontaneous symmetry breaking). The string tension or energy per unit string length is $\sim{\mu^2}$, where $\mu\sim{T^*}$ are the SSB scale and the temperature $T^*$ of the corresponding phase transition. The energy density due to these strings at the time of SBB is: \begin{equation}\rho ( string )\approx {\mu^2 l_h/{l_{h}^{3}}}=\mu^2/{l_{h}^{2}}.\end{equation} If the closed strings behave as CDM ( cold dark matter) particles then their mass density  $\rho( string)$ scales as $T^3$ and $\rho(string)\le{\rho(CDM)}$ should hold at any temperature $T$.The CDM density is: $\sim{5\rho(Baryon)}=5\eta_B n_{\gamma} 1 GeV$ with $\eta(B) \sim {5\times 10^{-10}}$ the baryon asymmetry. Thus
\begin{equation}
n_{\gamma} =N(dof)\zeta(3)/{\pi^2}T^{3}\approx{0.12 N(dof)T^3}
\end{equation}
with $N(dof)$ the number of degrees of freedom of relativistic particles at the temperature T. For $T\ge{TeV}$ we take $N(dof)_T=N(dof_{standard model})\sim{60}$ and the comoving dark matter density is :
\begin{equation}
\rho(CDM)= 2.5\times10^{-9}.60 T^3.GeV=1.5\times 10^{-7} T^{3} GeV
\end{equation}
To estimate $\rho_{string}$ at the time of the SSB  we need the horizon size $l_h(SSB)$ at SSB. In the radiation dominated era between the SSB and nucleosynthesis at time $ t=t_0\sim{1 sec}$ and temperature $T\sim{1 MeV}$, the horizon grows as $1/{T^2}$. Thus we have : $l_h(SSB) = l(T=MeV). MeV^2/{T(SSB)^2}= l(T=MeV) 10^{-6}/{\mu^2}$ with $l =ct_0\sim 3\times10^{10} cm$ and $T(SSB) \sim{\mu}$  the temperature or the scale of the SSB in GeV.  Thus we find $\rho_{string}\sim {\mu^6 10^{12}}$. Demanding that the strings will not saturate the CDM at $T=T(SSB)\sim{\mu}$ then yields the bound: 
\begin{equation}
\mu\le{10^7} GeV
\end{equation}
If the individual strings extending over the horizon stretch with the horizon growth the constraint obtained is much stronger. Much tighter bounds also follow from more careful considerations of the CMB fluctuations and unisotropy due the the strings - and when applicable also from em signatures.

The last and most extended type of topological defects are domain walls which result from spontaneous breaking of discrete symmetries. A pertinent example is the  left  right symmetry.  At  high energies an exact $SU(2)_LX SU(2)_R$ with mass-less $W_L , W_R$ gauge bosons with equal gauge couplings $g_L=g_R=g_W$ and a symmetric coupling to the Higgs sector is assumed.~\cite{Mohapatra, Senjanovic}.  Constraints from weak decays imply that the mass:$W_R\sim{g_W.\mu}$ is bound from bellow by  $2-3TeV$ and therefore the symmetry breaking scale should be $\mu > {10 TeV}$.  The initial $Z(2)$ symmetry of the lagrangian implies that we are equally likely to have regions where $m(W_L) \le{M(W_R)}$ where the important weak interactions al low energies are left handed and domains with the opposite, $m(W_R) \le{M(W_L)}$, pattern. The energy per unit area of the domain walls separating these region is $\sim{\mu^3}$. The energy $W=\mu^3 R_{Hubble}^2$ of just \textit{one} such domain stretching across the present horizon vastly exceeds by $10^{32}$  the total DM mass. In order to avoid the difficulty one might add a tiny potential breaking Z(2), $V=\epsilon \mu^4$ to push the domain wall away. This will introduce however a cosmological constant of order $\epsilon (10TeV)^4$, exceeding the ``observed" value of $[10^{-2}e.V]^4$ implying that $\epsilon\le{10^{-60}}$.

This fine-tuning problem replicates the fine tuning of the ordinary cosmological constant .The possibility that these two difficulties will cancel curing each other seems remote - but should be looked into. Because of frustration at junctures of more than two domain walls this remedy is not available for the case of SSB of $Z_N,\ N\ge 2$

\section{Implications of high reheat temperature for massive stable particles.}Many particle physics models have new massive particles $X$ with lifetimes $\tau(X)\ge 1 s$. This is clearly the case for massive particles which are CDM candidates as they should live much longer than Hubble time.  Indeed new physics at higher energy scales is often associated with new gauge interactions and/or new symmetries, suggesting that the lightest particle carrying the new Quantum number(s) be stable. A well known example is the LSP ( lightest Super-symmetric partner), stabilized by $R$ parity making it a good dark matter candidate - provided it leaves the correct relic density. Other frameworks  aiming to dynamically generate the E.W (electroweak) SSB via a fermionic condensate in a new confining gauge theory at a $\sim{TeV}$ scale also have massive long lived particles, e.g. the lightest baryons made of the new fermions.

A repeatedly encountered problem is that the annihilation rate of such massive particles is too small and may not reduce their relic density $ \rho(X) _{freeze-out}$ to (or bellow) the required $\rho(CDM)$. Matter -antimatter asymmetry, analog to our baryon asymmetry, can only increase the residual $\rho(X)$.\\

This difficulty is often evaded by assuming reheat temperatures lower than $M(X)$ or by having an additional massive particle tuned to decay at the right time so as to dilute the $X/{\gamma}$ ratio. The underlying assumption of this paper excludes these options. Future deeper insight into the origins of inflation may eventually justify it or point to specific cosmological scenarios where these remedies do naturally arise and are not introduced ad-hock in order to make certain pet models viable.\\

We next estimate of the freeze-out density recalling the ``WIMP miracle" that \textit{if} $M(X)\sim{TeV}$, then the right relic freeze-out density obtains for charge symmetric CDM annihilating into the lighter SM (standard model) particles with weak interaction cross-sections:
\begin{equation}
\sigma_{an}\sim{1/{v} (\alpha(W)/{M(X)})^{2}}
\end{equation}
In the above equation $\alpha(Weak)=g_W^2/{4\pi}=1/{30}$ is the coupling strength of weak interaction and $v=v(X)-v(\bar{X})= 2v(X)= 2k/(M(X) $ is the relative velocity of the $\bar{X}$ and $X$ in their center mass frame where $k$ is the common magnitude of the momenta. Since the scale of SUSY (super-symmetry) breaking and therefor the mass of the LSP's ( lightest SUSY partners) identified with X, were supposed be lighter than a TeV  to ``naturally" explain the mass scale of weak interaction, this gave further impetus to direct and indirect WIMP searches in underground detectors, in satellites, in ICE-CUBE and for their production along with all the other SUSY partners at the LHC (large hadron collider). The lack of positive evidence kept pushing up the masses of the WIMPS, the miracle stopped operating and over closure difficulties were encountered.

The number density $n(X)=n(\bar{X})=n$ of $X$ and $\bar{X}$ particles which are in thermal equilibrium with the rest of radiation at temperatures $ T\ge{M(X)}$, is exponentially suppressed once T falls below $M(X)$: $n\sim(\exp{-M(X)/T})$.  At a temperature $T_{f.o.}= M(X)/f$ the annihilation rate:
\begin{equation}
\Gamma(an) =nv\sigma_{an}\sim{n \alpha^{2} (Weak)/M(X))^{2}}
\end{equation}
falls bellow the Hubble volume expansion rate:
\begin{equation} 
\frac{\dot{v}}{v}=3H=3(8\pi/3\rho(T))^{1/2}/{M_{Planck}}.
\end{equation}
The $X$ particles then cease to be in chemical equilibrium and their co-moving density freezes out. The density $\rho(T)$ is that appropriate for radiation dominated era at temperature $T$ namely: $\rho(T)=N(dof)\pi^2T^4/{15}$. With $N(dof)$ the effective number of degrees of freedom at this temperature which we again take as  $N(dof)\sim{60}$ Using this in $\Gamma(ann)=3H$ with $T=T_{f.o}= M(X)/f$ yields $n(X)_{f.o}=2\pi/{5^{1/2}}m(X)^4/{m_{Pl}f^2\alpha^2}$. Demanding that the relic $X$ density together with the equal density of $\bar{X}$: $\rho(X+\bar{X})=2n(X) M(X) $ does not exceed the total allowed density of cold dark matter  $\rho(CDM)\ $ in Sec II above, yields an upper bound on $M(X)$:
\begin{equation}  
M(X)\le 100 TeV \alpha (1/f )^{-1/2}N (d.o.f.)^{1/2}(40\zeta(3)/{6 \pi^3.(8\pi/3)^{1/2}})^{1/2} 
\end{equation}   

Detailed calculations by many authors yielded $f$ values in the $20-30$ range. Taking $N_{d.o.f}$ to be $60$ and $\alpha\sim{1/{30}}$ , we find that unless $M(X)\le{TeV}$ we have excessive relic X density.  In general the bound on the mass scales as $(\sigma (ann) v \Omega(X) h^2)^{-1/2}$. The maximal annihilation rate used by GK $v.\sigma= \pi/{k^2}$ with $k^2 \sim{2T M(X)} =2M(X)^2 /f$ is bigger than the ``Weak value" featuring in the Wimp miracle above  by $\pi f 1/{\alpha^2}$ explaining their higher ``axiomatic" bound
$M(X) < 340 TeV$. In the following we argue that the maximal reasonable rate $v\sigma$ is  fact proportional to $1/{M(X)^2}$ rather than to $1/{k^2}$ and thus their bound is tightened not only by the obvious $1/{10^{1/2}}\sim 1/3$ due to $\Omega$ and $h^2$ decreasing since the time when the original GK paper was written by 5 and 2 respectively, but also by another factor of  $f^{-1/2}= 1/(5.5)$ so that the likely bound is: $M(X)\le{20 TeV}$. 

Absent precise statements for the non-diagonal annihilation amplitudes GK used rigorous bounds on the elastic $\bar{X}X$ elastic amplitudes which the optical theorem relates to total cross-section. However the qualitative difference between the t channel exchanges in elastic and annihilation processes ,which GK mentioned, suggests tighter upper bounds on the annihilation cross-sections lowering the upper bounds on $M(X)$ . We will also show how when the S matrix approach underlying the bounds breaks down, as it necessarily does for non-abelian gauge interactions with long range confining force, the limits are largely relaxed.

\subsection{When can large \textit{annihilation}  cross-sections arise?} 

This can be viewed from the s and t channels points of view. 

(i) s channel prospective

Enhanced cross-sections arise when the center mass energy $W(\bar{X} X)$ is close to $M(R)$ the mass of the particle $R$ which in turn is near the threshold: $M(R) = 2 M(X)+m$ with $m/M$ $\le\le 1 $. For negative $m$ the particle $R$ manifests as a pole in the $\bar{X}X\rightarrow i\bar{i}$ amplitudes $A(X,i)=r_{X,i}/(W-M(R))$. For positive m it is complex pole at  $W=M(R) +i\Gamma/2$ with $\Gamma$ the total width of the corresponding resonance. This width is the sum of the elastic $\bar{X}X$ width and the widths $\Gamma_{i}$ for R to decay into the final $(\bar{i} i)$ states.  

\begin{equation}
\Gamma=\Gamma(el)+\sum_{i}{\Gamma_i}. \textrm{ Total width is sum of partial widths}
\end{equation} 

The residue of the pole in the annihilation amplitude: $r_{X,i}=g(\bar{X}X;R). g(\bar{i}i;R)$ factorizes into the product of the couplings of R to the initial and final particles. (For simplicity we assumed final states with two particles. These particles could be unstable and decay later.) (see Fig 1a)

The pole or the resonance contribute to a particular partial wave with $J =J_R$, the angular momentum of the particle R. At  freeze-out the temperature $T$ is $T_{f.o}=M_{R}/f \sim M(R)/{30}$. The $\bar{X}X$ system of interest is then non-relativistic and $J=L+S$ with $S$ the total spin of $\bar{X} X$ and $L$ the orbital angular momentum in the Lorentz frame of the their center of mass. The partial wave expansion for the $\bar{X} X$ elastic scattering amplitude $f(\theta)$ in its familiar non-relativistic form is: 
\begin{equation}
f(\theta,W)=\sum_l{(2l+1) a_l(W)/{ik} P_l[cos(\theta)]}; l=0,1,....\infty  \textrm{ with } \left ( d(\sigma)/{d(\Omega}\sim f^2(\theta) \right )
\end{equation}
where $P_l [cos(\theta)]$ are the Legendre polynomials and $k=(W^2-4M(X)^2)^{1/2}/2$ is the momentum of $X$ or $\bar{X}$ in their center mass frame. In writing the above expansion we implicitly assumes spin-less particles so the J=L and we do not have to separately expand the 5 helicity amplitudes for say $1/2+1/2\rightarrow{1/2+1/2}$ or ${1+1}$ process where spin half X's annihilate into electron positron or $W^{+}W^{-}$ pairs. The arguments do however carry over with minimal changes to these cases as well.

Unitarity limits the magnitude of each partial wave amplitude by $a_l(W)\le{1}$ . Thus the maximal contribution of any partial wave to the total cross-section is $(2L+1) \pi/{k^2}=\sigma(max)_L $.  This maximal value is achieved only at  $W=W_R=M(R)$ , the peak of the the Breit Wigner distribution corresponding to the resonance $R$:
\begin{equation}
\sigma(\bar{X} X\rightarrow{\bar{i}i})=(2J_{L}+1) \pi/{k^2} \Gamma(el)\Gamma(i)/{[(W-M(R))^2+\Gamma^2]}\label{Breit Wigner}
\end{equation}The annihilation cross-section into any $\bar{i}i$ channel is proportional to the product of the decay widths $\Gamma(X)=\gamma(R\rightarrow(\bar{X}X)$ and $\Gamma(i)=\gamma(R\rightarrow(\bar{i}i))$. Since $\Gamma(X)\sim{g(\bar{X}X;R)^2}$ and $\Gamma(i)\sim{g(\bar{i}i;R)^2}$ the residue at the complex poles factorizes.

Thus three factors jointly generate a large annihilation cross-section. To saturate the KM bounds over an extended energy range all three factors should be maximal.

The first is the $2\pi/{k^2}=\pi/{M(X)K(X)}$ factor with $k^2/{M(X)=K(X)}$ - the kinetic energy in the center-mass system of the $X$ particles. This kinetic energy \textit{cannot} be arbitrarily small. The temperature during the relevant period of freeze-out is: $T_{f.0}=M(X)/f$ and since the X particles are in thermal equilibrium $K(X)\sim {T}\sim {M(X)/f} $. Thus this first factor can enhance the annihilation cross-section by $1/f \sim 3$ relative to  $2\pi/{M(X)^2}$. 

The second factor is the full width $\Gamma$. The rate of annihilation is :
\begin{equation}
\int dk(1)dk(2) n(X) [ k(1)] n(\bar{X})[k(2)]\sigma_{ann}(W(1,2)) v(1,2)
\end{equation}
with $n(X)$ and $n(\bar{X})$ the number densities of the $X$ and $\bar{X}$ particles with momenta $k(1)$ and $k(2)$ and W(1,2), v(1,2) the total energy and relative velocity in their center mass frame. In the thermal environment $W(1,2)$ is spread over $\Delta(W)=T$. For the annihilation cross-section $\sigma_{ann}(W(1,2))$ to be large over the full energy range the resonance R has to be sufficiently broad: $\Gamma\sim{\Delta(W)\sim(T)}$. Using $T=T_{f.o}= M(X)/f $ the condition becomes: $\Gamma > {M(X)/f}$. Broad resonances naturally arise if  $\bar{X}$ and $X$ strongly interact with each other though a near threshold bound state pole or resonance requires tuning. The total width is the sum of the width to decay back to the original  $\bar{X} X$ and the decay widths into the annihilation channels $R\rightarrow {\bar{i} i}$. The first width  $\gamma (R)\rightarrow{\bar{X}X} =g(R,\bar{X} X)^2 k/{8\pi}$ is  suppressed by $k/{M(X)}=v(X)={f^{-1/2}}\sim{1/5}$. 

The third ingredient needed to ensure large annihilation cross-section are large decay widths $\Gamma(i)=g^2(R ; \bar{i}i) M(X)$. These decay widths are not suppressed by kinematics and jointly can contribute half the the total width maximizing annihilation cross-section. However we claim that in most natural scenarios these are strongly suppressed dynamically a feature best seen from the t channel prospective (ii). 

(ii)  t channel induced large cross-section and their limitation in the case of annihilation.

Exchanging mass-less particles in the t channel generates infinite elastic scattering. This follows from the infinite range of the corresponding potential and the Rutherford scattering via photon exchange is a prime example. Indeed  while unitarity limits the cross-section contributed by each individual partial wave the sum over the partial waves: 
$A(s,t) =A(s;\cos(\theta)= \sum(2L+1) a_{L}(s) P_{L} ( \cos(\theta) )$ with $\theta$ the center of mass scattering angle could diverge. 
The invariants made up of squares of the sums of four-momenta entering in the different channels : $s=(p(X+p(\bar{X}))^2 =W^2, t=(p(X)-p(i))^2$ and $u=((p(X)-p(\bar{i}))^2 $ are often used in discussing $2\rightarrow{2}$ amplitudes. For finite mass $\mu$ of the exchanged particle the elastic amplitude $A(s;cos(\theta))$ is analytic in the ``Lehman Martin" ellipse in the $cos (\theta)$ plane with foci at  $+1,-1$ and a semi-major axis $1 + t_0/{2k^2}$ with $t_0=\mu^2$ the nearest t channel singularity and the sum converges therein. 

A detailed analysis yields an exponential decrease of the higher partial waves $a_{j}(s)\approx (k/{\mu})^{2l}$
\footnote{For completeness we briefly sketch the proof of this.  A t-channel exchange of a particle with mass $\mu$ contributes to the scattering amplitude $f(k,x)$ a term $1/{(\mu^2+t)}=1/{(\mu^2+2k^2(1-x))}=\mu^{-2}\times1/{(2k^2(1-x)/{\mu^2}+1)}$ with $x=\cos(\theta )$ and $\int_{-1}^{1}{P_l(x) \mu^{-2}\times 1/(2(k^{2}(1-x)/(\mu^2)+1)}$ is the corresponding $l$th partial wave. For $k^{2}\ll\mu^{2}$, we expand the integrand in a convergent geometric series: $\sum_{n} (-1)^{n} (2 k^{2} (1-x)/\mu)^{2n}$. The completeness and orthogonality properties of the $P_l(x)$ imply that the $n=l$ term is the first making a non-vanishing contribution to $a_l(k)$, which is actually: $(2k^{2}/(\mu^{2})^{l} \times I(l)$ with $I(l)=2\int_{0}^{1} P_l(x) x^{l}$.  Substituting the Rodrigez formula: $P_l(x)= 1/(l!2^l) d^n/((dx)^n)(x^2-1)^l$ and integrating $l$ times by part and canceling the $l!$ we find that $I(l)=2^{-l}\int_{-1}^{1}(x+1)^{l} (1-x)^{l} \le 2^{-l}$. Finally, we have $a_{l}(k)\le {(k^2/(\mu^2)^l}$ and $\sigma_l(k) \le (k^2/{\mu^2})^{2l}$.}. The extra velocity factor suppression of the p annihilation due to the p wave required for Majorana relics is well known. Still the elastic cross-section $\sim{\mu^{-2}}$ are large for small $\mu$. However, the $\bar{X}X\rightarrow \bar{i}i$ annihilation cross-sections of interest have \textit{no} t channel enhancement. This follows from both kinematics and dynamics. We are interested in annihilation near threshold where $W=s^{1/2}\sim 2M(X)$. At this point $k\approx 0$ and $t\approx u$.  Recalling that $s+t+u= 2M(X)^2+2 m(i)^2\sim{2M(X)^2}$ and $s\approx 4M(X)^2$, we find that $t\approx -M(X)^2$. Thus, the annihilation amplitude is suppressed by $1/{t(min)+\mu^2}\le{1/{M(X)^2}}$.
The annihilation cross-section is in fact smaller by another factor $4$. The reason is that the lightest particle that can be exchanged has a mass $\mu\approx{M(X)}$ (see Fig. 1b) as otherwise the $X$ quickly decays into the final i and the exchanged $\mu$ particles and would not be long-lived contrary to our assumption. Note that the annihilation: $\bar{X}X\rightarrow{\bar{X'}X'}$ where $X'$ is just slightly lighter than $X$ and with a light  $\mu$ exchanged in the $t$  channel and a large cross-section: $\sigma\sim{1/\left ( \mu^{2}+(M(X)-M(X')^2) \right ) }$ is also excluded. In this case the heavy $X' $ is the lighter particle in the new sector and therefore would be stable. Another X'' will have then to be invoked to have the $X' $ annihilate quickly etc. More importantly all non zero l terms in the partial wave expansion of the annihilation cross-section are - because of the short range t channel induced interactions  $r_0\approx{ 1/{M(X)}}$- drastically suppressed at threshold by the factor of $(k/{M(X)})^{2l}$. This makes the contribution of all none-zero $l$ states negligible leaving us only with 
$J^P(R) = 1^-$ or $ 0^-$ for spin $1/2 ,X$ particles or with $J^P(R)=0^+$ for spin zero $X'$s and we cannot use $t$ channel exchanges to enhance the rate of annihilation.

All of this just elaborates on and accords with KM. There are however several new points which we expound next.

(1) The t channel exchanges can provide a potential in the $\bar{X}X$ channel tuned to produce a near threshold $\bar{X}X$ resonance/ bound state, in the l=0 partial wave. However we still \textit{cannot}  saturate the upper bound $\sigma_{ann}=\pi/{k^2}$ in this partial wave and annihilation cross-section enhanced by $(M(X)/k)^2=f $ relative to the nominal $1/{M(X)^2}$ are \textit{not} obtained.

We argued above that annihilation requires that a particle of mass $\sim{M(X)}$ be exchanged in the t channel and the range of the ``annihilation potential"  $r_{0}=1/M(X)$, which is definitely smaller than the effective wavelength size $1/k$ of the resonance (See Fig 3). This suggests a geometric reduction of the cross-section by $k^2/{M(X)^2}=1/f$ which certainly would hold at high energies, It is however also the case at low near threshold energies as the following argument indicates.  During the lifetime $\tau(R)=1/{\Gamma}=f/M(X)$ of the resonance $R$, the $\bar {X}$, say, travels a distance of $d=v \tau$ relative to X which we assume to be fixed at the origin. Using $v=k/M(X)=f^{-1/2}$ we find that $d =f^{1/2}/M(X)$ and the effective annihilation volume swept by the $ \bar{X}$ is $Vol(ann)=\pi r^{2}_{0}d= \pi f^{1/2}/{M(X)^3}$. The full  volume  of the resonance (or the corresponding bound state obtained by $k\rightarrow ik$)  is $Vol(tot)=1/k^{3}= f^{3/2} M(X)^{-3}$  namely f times larger. The probability of actual annihilation rather than scattering back into the initial $\bar{X}X$ state is then the ratio of these volumes namely $1/f$ . 

Thus we are back to the original $\sigma(ann)\sim{1/{M(X)^2}}$ estimate and upper bounds on M(X) larger by $f^{1/2}\sim{5}$ ensue. 

It is important to note that the large annihilation cross-section of anti-proton and proton near threshold does \textit{not} contradict the above arguments for suppression of $\bar{X}-X$ annihilation. There are several important differences between the two cases. Indeed $p-\bar{p}$ ``annihilation" is a misnomer. Unlike in the genuine 
$e^+ e^{-}\rightarrow {2\gamma}$ annihilation where the electron and positron disappear no quarks and anti-quarks need to be annihilated in $\bar{p} p$ annihilation. Rather the initial  $q^{3} + \bar{q}^{3}$ state rearranges into a three meson $(\bar{q} q)^3$ configuration.  As the slow $p$ and $\bar{p}$ approach each other near threshold, the levels of these two configurations cross and there is an almost complete adiabatic transition of $\bar{p}-p$ into the three mesons. Since no heavy $t$ channel exchanges are involved the only suppression of this process comes from the finite nucleon size. This issue will be important  for our subsequent discussion of the relic abundance of heavy $X$ particles which carry the charges of a confining gauge theory and generate stable  or long lived baryons. 

There is however a clear distinction between this and what is rigorously excluded in the ordinary S matrix formalism -namely the cases in which the GK bound is violated. Disregarding any t channel dynamics we can postulate a new heavy gauge particle or a scalar of a mass M(R) tuned to be within a factor of $1/(2f)$ near the $2 M(X)$ threshold. We then arbitrarily prescribe that it's coupling to both the $X \bar{X}$ initial particles and the final $i\bar{i}$ particles so as to saturate the GK bound. (see Fig. 2)

At this point we depart from the purely formal approach which does allows particles as heavy as $\approx{100 TeV}$ and go beyond the original discussion of  GK.  Utilizing the many experimental bounds on WIMPS which have accumulated in the meantime and also some general comments on the underlying theory we discuss the potential difficulties entailed if   new particles of the type described above are allowed . 

The light particles into which the $R$ decays can be new 'dark sector' $x$ particles. Since the lightest particle in this sector is stable, potential over-closure still looms and one needs detailed scenarios where the new $x$ particles are sufficiently light to meet CDM or even WDM or HDM ( warm dark matter and hot dark matter) requirements.

The other alternative- that the R decays into SM particles is more interesting but more restricted phenomenologically. If the possible $i\bar{i}$ final states include the u and d quarks then the maximal couplings of R to the $\bar{X}X$ dark matter particles and to quarks/nuclei can generate excessive X-nuclear cross-sections. With target nucleus mass  M(A,Z) these cross-sections are  $\sigma(X-(A,Z))\sim{1/(M(X)^4). M(X).M(A,Z)} =4.10^{-35}cm^2.(M(X)/{TeV})^{-3} A/{100}$ and may exceed the extremely tight bounds on nuclear cross-sections inferred from direct searches. On the more theoretical side introducing a new \textit{strong} gauge interaction common to the SM and the dark sectors which would be implied if $R$ is a vector particle seems to be a nontrivial undertaking, even if it is only a $U(1)$ interaction. This is also the case if $R$ is a scalar. Having a multi-TeV elementary scalar field couple to the electron or muon say with O(1) couplings when the only scalar couplings of the latter to the Higgs particle are of order $10^{-6}-10^{-4}$ seems awkward. The new scalar or  vector interactions may also modify precisely tested flavor physics in rare decays of standard model particles. An interesting possibility is that the new dark sector has it's own rather light photon which we denote by $A'$ corresponding to some new $U'(1)$. In this case one could  have the annihilation $\bar X-X\rightarrow{A' A'}$.

Since an X exchange is involved, the cross-section then has the generic $\alpha'^{2}/{M(X)^2}$ form. The same argument applies for annihilation via either t channel or s channel exchange to standard model particles. Specifically in the s channel the R resonance needs to mix with some standard model neutral particle. These are much lighter and will be off-shell by the large amount of $\sim{(M(R))^2}$. \footnote {An example from hadron physics is the $\rho$ resonance at $760 MeV$ (playing the role of R) in the $\pi \pi$ channel, the analogs of the $\bar{X}X$ particles here. While the $160 MeV$ broad resonance dominates $\pi \pi$ scattering for $W=600-900 MeV$ the annihilation into the light $e^{+}e^{-}$ pair has a branching of $\sim {10^{-5}}$ due to the weak em coupling and the need to exchange a virtual photon in the s channel which is off shell by $m(\rho)^2$.}

\section{IV. Can the upper bound on $\sigma_{annihilation}(\bar{X}-X)$ and resulting bounds on M(X) be relaxed.}

The lengthy discussion above suggested that the only way to enhance the annihilation cross-section beyond the bench-mark of $\alpha^{2}/{M(X)^2}$ is to tune parameters so as to have a near threshold $\bar{X} X$ resonance or a near zero energy S wave bound state. Even then because the temperatures involved are $T_{freezout}=M(X)/f \approx {(0.03)M(X)}$ the effective cross-section determining the relic freeze-out density $n(X)_{f.o}$ is  only $\pi/{k^2}=\pi/{M.T}\sim{\pi/{MT_{f.o}}}$. We have further argued  that even the resulting $f = {M(X)^2}/{k^2}$ enhancement relative to $\sigma(ann)=1/{M(X)^2}$ is absent if the process proceeds via t channel exchange of particles as heavy as $M(X)$ . This along with $10^{-1}$ decrease of the allowed relic density tightened the upper bound by $(10/f)^{-1/2}\sim 1/{17}  \le{TeV}$ from $M(X) \le{340 TeV}$ to $M(X) \le{20 TeV}$.

There are many beyond standard models with stable or long lived X particles more massive than this limit.~\cite{Esmaili, Bhattacharya, Bella}\footnote{one example among many is an X particle with $M(X)\sim{5 PeV}$ and lifetime of $10^{28} s$ the decays of which generate the Ice-Cube PeV neutrinos.~\cite{Esmaili, Bhattacharya, Bella}}.  If we exclude cosmological scenarios with a reheat temperature lower than M(X) or with a late inflation or late decays of other heavy particle dilutes the X density we need to modify the X dynamics in order to enhance the annihilation rates. We next discuss the possibility that new long range interactions not included in the standard $S$ matrix framework can strongly enhance the annihilation cross-sections and relax the upper bounds on $M(X)$.

It is well known that the Coulomb like interaction due to the exchange of a relatively light new ``dark photon" $A'$ of a new $U'(1)$ of coupling to $\alpha'$ and mass $m(A' ) \le M(X) \alpha'$ pulls the $\bar{X }$ and $X$ closer together enhancing the wave function at zero $|R(X) -R(\bar{X}|$ separation and the ensuing annihilation rate via some short range exchange in the t channel. In this case ( but \textit{not} for non-abelian interactions), a kinetic mixing $\epsilon F.F'$ with the ordinary photon is possible (providing the dark photon is not mass-less). The resulting richer scenarios allow also $X+(A,Z) \rightarrow{X+(A,Z)}$ scattering in direct detection searches for D.M and $A'$
production and subsequent ,say $A'\rightarrow{e^+e^-}$, decays. These scenarios were extensively studied~\cite{ArkaniHamed1, Bjorken}, and the role of the Sommerfeld enhancement (SE) of the $\bar{X}-X $ present annihilation rates was emphasized.The resulting Sommerfeld enhancement of the annihilation rate is by $\alpha'/{v(X)}= SE.$~\cite{ArkaniHamed2} 

The S.E enhancement occurs at  temperatures $T$  smaller then the Coulomb binding $BE=(\alpha')^{2} M(X)/4$. However even if $T/BE$ is $\approx 1/{100}$ and $v\approx{1/{10}}$ we find even for $\alpha' \approx 1$ that $S.E.=\alpha'/v =10$. The upper bound on $M(X)$ is then further relaxed only by $S.E^{1/2}\sim{3.1}$.  Does the above moderate S.E. fully exhaust the effect on the $\bar{X}-X$ annihilation of the light $A'$. In principle $ \bar{X}-X$ bound states can form via the : $\bar{X}-X\rightarrow{(\bar{X}-X)_n +\gamma'}$ analog of ordinary recombination $e^{-}+p\rightarrow H_{n}+\gamma$. The subscript refers to the n'th excited state with binding $E_n=Ry'/{n^2}$ and size $r_n=r_{Bohr}'.n^2$, where $ Ry'= M(x).\alpha'^2/4$ and $r_{Bohr}'=2/{M(X)\alpha'}$. Once in the ground state the local effective density becomes$n(X)_{effective}\approx{ar_{Bohr'}^{-3}}$ which can be as much as $10^{27}$ bigger than the low cosmological density and annihilation will proceed immediately. The recombination cross-section for capture into the n'th state is~\cite{ArkaniHamed2, Bethe} $\sigma_n= C.n/{\alpha' M(X)^2}$ with C a constant of order unity.  However, capture into the high n states cannot enhance the overall $\bar{X}X$ annihilation since it occurs only at temperatures $T_n\le M(X) \alpha'^2/{n^2}$ where the $X$ number density $n(X)\sim{T^3}\le n^{-6}$ becomes too small for the rate of recombination to keep up with that of the Hubble expansion which scales as $T^2$. Also, unlike ordinary recombination where $n(proton)/{n(\gamma)=n(e)/{n(\gamma)}}$ is a constant value fixed by the baryon asymmetry, the ratio of $n(X)/{n(\gamma)}$ keeps decreasing so as to become smaller than $\frac{n(Baryon)/{n(\gamma) 5m(p)}}{M(X)}$ where the factor $5$ is the ratio $\Omega(DM)/{\Omega(baryon)}$.

Much stronger enhancements of the annihilation cross-section arise if the $X$ particles have non-abelian confining interactions at relatively low scales. In the simplest such scenario the heavy X particles carry ordinary $SU(3)_c$ color as was the case for the gluino $\tilde{g}$ in split SUSY~\cite{ArkaniHamed2, Bethe}. At the QCD phase transition with temperature $T=T_{c} \approx \Lambda(QCD)\approx{200} MeV$ the $g^{~s}$ may pair into color singlet and annihilate but are more likely to bind to the abundant ordinary gluons to form $\tilde{g}-g$ color singlet glue-ballinos. The size of this state is fixed now by the QCD scale to be $\sim{0.3\ fm}$ and the cross-section for collisions of  these particles is far bigger than that of the direct $\tilde{g} \tilde{g}$ annihilation. Further, in a substantial fraction of the collisions the glue-ballinos rearrange into a deeply bound $\tilde{g}-\tilde{g}$ gluino ball and a light $gg$ glue-ball with the subsequent quick annihilation of the gluino-ball. The net effect is to revive the $\bar{X}-X$ annihilation later than and over and above the early annihilation which  froze out and a much higher $T\approx{M(X)/{30}}$. Various estimates suggest that this indeed allows for relatively stable gluinos at masses higher than the above bounds. If the gluino was the lightest among all the SUSY partners and therefore stable (which in most models it is not) and further the above annihilation left just the right amount of relic $g\tilde{g}$ particles to be dark matter these could indeed be the dark matter particles. However just the residual gluon exchange interactions between the R particles and ordinary hadrons would cause them to bind to nuclei forming ultra heavy isotopes. The extremely strong bounds on such isotopes then exclude this and many other SIMP scenarios.

These last difficulties are evaded if the X particles carry another color' and be the ``quarks", $Q'$ of another say $SU'(N)$ non-abelian gauge interactions. The confinement scale associated with  the new group has to satisfy : $M(Q')\ge  \Lambda'$. Otherwise the Q's  could confine into $Q'^{N'}$ color' singlet  baryons of a large mass $\sim {N'\Lambda'}$. These stable baryons will have a small size  $R(Q')\sim {1/{\Lambda'}}$ and correspondingly small annihilation cross-sections $\sim{\Lambda'^{-2}}$ and the difficulty encountered with the original $X=Q'$ will repeat all over again. The model leads to two different scenarios depending on if  q's of mass $m(q')\le{\Lambda'}$ transforming  as the fundamental N' representation of $SU(N')$ exist or do not exist.

The first case is analog to QCD with light ud quarks. At the $ T=\Lambda' $ mesons $M= Q'\bar{q'}$ and $\bar{M}$ form. $M \bar{M}$ collide with large ($\sim{\Lambda'^{-2}}$) cross-sections and in $\sim{1/2}$ the cases rearrange into $\bar{Q'} Q'  + \bar{q'} q'$ and the $Q' \bar{Q'}$ inside the heavy quarkonia will then shortly annihilate.  Note however that the heavy Q's might the just like the t, b and c quarks in QCD mix with and decay into the lighter q's and the longevity of the Q's will then be compromised.

These difficulties are avoided if the Q's are the lightest particles carrying the SU'(N) quantum numbers rendering them completely stable. The confining phase transition at $T^{*}=\Lambda'$ leads to complete $\bar{Q'}-Q'$ annihilation by long string like chromo-electric flux tubes connecting the residual $bar {Q'}Q'$ remaining after the  earlier freeze-out of the perturbative annihilation at a temperature $T_{f.o} \ge \ge T^{*}$. These strings then relax to their small ground state and the $Q'-\bar{Q}'$ annihilate. We need however to address the question of $Q'^{N}$ baryons. We note that the formation of  baryon- anti-baryon pairs is suppressed relative to the dominant $\bar{Q}-Q$ formation by an essentially combinatorial factor of $\sim{\exp{-N}}$ due to  the need to separate the $N$ Quarks and the $N$ anti-quarks forming the baryon and the anti-baryon respectively. This is very much analog to the $\exp{-1/{\alpha}}$ factor suppression of monopole pair production. The latter is also suggested by the fact that large N baryons can be viewed as topological skyrmions with the large baryon charge being topological like the $ \approx{1/{\alpha}} $ magnetic charge of the monopole.~\cite{Witten}. It seems unlikely that the annihilation of large N baryons will be suppressed by a similar exponential factor. The baryons do not really annihilate but only rearrange into N $\bar {Q}- Q$ pairs. For low enough velocity and overall attractive interaction the process is adiabatic (with many level crossings) and annihilation is virtually ensured once the baryons are within $1/{\Lambda}$ range.  Also that decays of the $X$ baryons into light say SM particles might also proceed via some GUT like exchanges. These will be faster than the proton decay in $10^38$ Sec by $(M(X)/{m(N)})^5$ factors which for $M(X) \ge{10^8 GeV}$ may yield acceptably short lifetimes.

The resulting avoidance of the rigorous S matrix upper bounds on the masses of relic stable particles is not surprising. The confining linear potentials between the $Q'$ and $\bar{Q}'$ have no S matrix counterpart where all potentials are superposition of Yukawa potentials with different ranges: $V(r)=\int d(\mu^2)\rho(\mu^2) \exp(-\mu r)/r$. The spectral function $\rho(\mu^2)$ corresponds to various exchanges in the t channel. In particular a mass-less exchanged photon generate to $1/r$ potential the longest range potential possible  and a linear potential cannot be achieved \footnote{The conflict between confinement and S matrix has been particularly emphasized by the late G. Preparata}.

At the $QCD$ phase transition all relevant length scales - the average separations of quarks or of gluons and the sizes of the forming $\bar{q} -q$ and $gg$ states happen to be the same: of order  $1/T\approx{1/{\Lambda}}$, and all $\bar{q}-q$ pairs which are spatially close readily transform into the confined meson states. Here, however, the density at $T=T'_c\approx{\Lambda'}$ of the heavy $Q'$ which survived earlier annihilation is smaller by about at least 15 orders of magnitude than that of the the g's ( the SU(3)' gluons) $n(g')\sim{T'^3} \approx {T^3}$ and the confinement of the $\bar{Q}Q'$ pairs seems rather puzzling.

Thus $|R(Q'_i)-R(\bar{Q'}_j)|$, the separation between near by pairs exceed the natural $1/{\Lambda'}$ scale by $10^4$ or more.  Prior to confinement in a Coulomb like phase the chromo-electric fields the  flux/field lines connect any given $\bar{Q}_i$ anti-particle to several $Q_j$ particles. At the phase transition this radically changes. The $Q'(i)\bar{Q}'(j)$ relation becomes ``monogamous" with  each $\bar{Q'}$ is paired with the nearest $Q'$ and all the flux lines emanating from it to other $Q's$ disengage , jump over and collapse to the long string connecting to this specific neighbor. This could not happen abruptly. Indeed if the phase transition is of first order that can be the time during which the system loses latent heat and the predominantly $B^{2}$ condensate gradually builds up starting with small bubbles which keep multiplying and expanding and in the process keep via the inverse Meisner effect pushing on the E flux line all the way until they form the final relatively thin and elongated strings and in particular assume the configuration where the total energy of the system: $\sim \Sigma|R(Q'_i)-R(\bar{Q'}_j|$ is minimal.\footnote{This intuitive picture was pointed out to me by G. Stermann}

Other heavy colored particles suggested by Luty are the Quirks which carry \textit{both ordinary and a new color}. The arbitrary yet fascinating aspect of the theory is that while the Quirks are much heavier than the ordinary quarks - the scale of the new confining gauge theory is assumed to be (much) lower than that of ordinary QCD.: $\Lambda' \le \Lambda (QCD)$, a feature that leads to new signatures in LHC.\cite{Kang1, Kang2} and to a very interesting cosmology.

We~\cite{Jacoby, Nussinov} have argued that the various late stages of annihilation at and after the color and color' phase transitions reduces the relic density of Quirks to be bellow that of CDM without invoking a reheat temperature lower than M(Quirk). Also an appropriate $\sigma'$ tension of the new color' strings, can suppress generation of super-heavy isotopes made of  Q's bound to heavy nuclei by the Quirks produced over earths  lifetime by energetic cosmic rays in earths atmosphere. We still need to ensure that the residual density of $g'g'$  glueballs which form at the time of the color' confinement phase transition at temperature $T=T'_c\approx {\Lambda'}$ would not leads to unacceptable residual density. This issue is discussed among oth8er cases of  over-closure due to weakly coupled light particles in the next section.

\section{Sec VI  bounds on light stable (possibly dark matter) particles.}

If we do not allow low reheat temperatures then also many scenarios with light x particles with masses $m(x)$ in the 0.1 KeV to MeV range, are excluded. Such light, weakly coupled particles have very long lifetimes against decay and or annihilation into standard model particles . Unless an appropriate decay/or annihilations into as yet lighter hidden sector particles is provided then these particles - which as we argue bellow have been in thermal equilibrium at some early epoch would have an excessive relic density and over-close the universe. The models harboring these particles will then be excluded simply because they cannot be assumed to be too weakly interacting so as to be produced only in out of equilibrium processes. 

The Quirk model mentioned above, supplies the first example. We need to ensure that the residual density of $g'g'$ glueballs which form at the time of the color' confinement phase transition at temperature $T=T'_c\approx {Lambda'}$ would not lead to unacceptable residual density. The decay of $g'g' \rightarrow 2\gamma$  can occur in the Quirk models with light ($\Lambda'\le \le \Lambda(QCD)$ only via a $Q'$ box followed by $q$ box diagrams. Because of gauge invariance these box diagrams generate Heisenberg-Euler type effective lagrangians color' involving the  QCD and e.m field strengths $F_{\mu,\nu}^2 g_{\mu,\nu}^2$. The resulting decay rate $\Gamma \sim{ \alpha(em)^2\alpha(QCD)^4\alpha'^2 M(Q')^{-8 }\lambda(QCD)^{-8} .m(g'b)^{17}}$ is way too small. Simple thermodynamics and conserved entropy imply that processes $3 gb \rightarrow {2 gb}$ first noted in~\cite{Carlsen} cannot, contrary to our earlier expectation , reduce $n(g'g')/{n(\gamma)}$. In the quenched approximation for $N'=3$, lattice calculations done for QCD and which are fully justified here with no light Quirks, suggest that $m(g'g')\sim {7-10 \Lambda{QCD}}$ during the prolonged first order phase transition at the temperature of $T'=\Lambda'$ where a lot of latent heat is emitted still into the gluonic gas we can expect some significant reduction by a factor of $\sim{\exp{-m(g'g')/{T'}}}\le\exp{-7}\sim {10^{-3}}$. The number reducing process we considered has been recently revived in~\cite{Hochberg} where a new paradigm for O$(40 MeV)$ dark matter has been introduced and where the extra entropy emitted by the dark matter particle reducing interactions, is dumped into the normal SM sector which for a long while is in thermal equilibrium with the dark sector.

Generically the interactions producing the x particles are suppressed by some heavy new physics scales $\Lambda_{new}$  and proceed with rates proportional to $T^{n+1}/({\Lambda_{new})^{n}}$ with $n =3, 4$ or $5$. Still for temperatures $T$ which are comparable or even some-what smaller than this new scale the x particles would have been in thermal equilibrium. A well known example is provided by the the ordinary left-handed neutrinos which are coupled to all the remaining SM species and are in thermal chemical equilibrium so long as the temperature exceeds O( MeV). This is because the relevant rate scales as $G_{Fermi}^{2}T^{5}\sim{T^5/{v^4}}$ with $v\sim 0.2 TeV$ the scale of ordinary weak interaction falls bellow the Hubble expansion rate $T^2/{M(Planck)}$ only at this low temperature. At the (at least!) $ \sim 10^{10}$ times higher reheat temperature envisioned any other even supper-weakly coupled species such as sterile neutrinos or ALPS ( axion like particles)  would have to be thermally produced. This in turn may exclude a very broad range of scenarios where long lived x particles more massive than say $100 eV$ are required.  In particular it would negate the possibility that  sterile neutrinos of mass $\sim{7 KeV}$ and lifetimes of $\approx{ 10^{27}}$ seconds generate the recently observed $3.5 KeV$ X ray line~\cite{Abazajian}. We should emphasize that this reasoning does not apply to \textit{all} forms of light long lived particles. An example is provided by the dark photon mentioned above. If there is no dark sector and the lightest particles carrying the new U'(1) are extremely almost Planck scale heavy than its may have never been in thermal equilibrium. Its only coupling is via the mixing $\epsilon$ with the ordinary  photon which for masses $m(A')\le{MeV}$ decay only via an electron box diagram into three photons and with lifetimes which are far longer than Hubble time. This is so because for such light dark photon considerations related to the cooling of supernovae by photon' emission strongly bind the strength of its mixing with the ordinary photon $\sim{\epsilon^2} \sim{10^{-20}}$. The cross-section for producing this dark photon e.g. $\sigma{\gamma+e\rightarrow\gamma'+e}\sim\epsilon^{2}\alpha^{2}/{T^2}$ yields then in the relevant radiation dominated era where $n(e)\sim{T^3}$ a rate of production $\Gamma\sim{\sigma n(e)}\sim T$. unlike the cases considered above $\Gamma$ increases more \textit{slowly} with temperature than the Hubble expansion $\sim T^2/{M(Planck)}$ and hence can be out of equilibrium even at the highest possible temperatures. Hence we do not encounter the over-closure difficulties and could have the dark photon be warm dark matter.

\section{Section V. Summary and conclusion}
This paper was inspired by the possibility that the BICEP data suggest a very high scale of the primordial inflation. \textit{If} we could infer from this that the reheat temperature is also rather high and \textit{if} there were no very significant later inflations, then many physics scenarios could be excluded and at least strongly constrained. Specifcally these are scenarios with topological defects and/ or heavy ($M(X)\ge{TeV}$ stable or long lived particles particles or even relatively light $m(x) \sim KeV$ particles.

It appears now that the early Bicep claims of a strong tensor part in the CMB and the need for a strong early emission of Gravitational waves may not survive. It is still worthwhile to describe the many implications of assuming a high reheat temperatures and  of absence of subsequent appreciable inflations. While most of the material presented is not new and was discussed in various specific cases we believe presenting them jointly and the few improvements we have made are of interest.

 The main part the paper dealt with the difficulties in scenarios with of heavy long lived particles.  Under the above assumption these particles were at an early stage when the temperature $T\ge{M(X)} $ in thermal equilibrium. The rather strong upper bounds on their annihilation cross-sections derived in standard S matrix approach translate into lower bounds on the residual freezeout density of these X particles. Unless they are lighter than $\approx {20 TeV}$ the residual density will exceed that of the allowed CDM.
 
There are several ways - beyond giving up our assumed cosmological inputs of large reheat temperatures and no diluting late inflation -to avoid these bounds we could  use the late decay of some other long lived massive particle to increases the entropy and dilute the number ( and energy) density of the relic X's . alternatively we have to ensure that the X particles decay fast enough an option which does not exist in the interesting case where the new particles constitute the dark matter.

We have  noted that new physics and new physics scales are often associated with long lived particles. This can be due to new quantum numbers associated with the new physics which are conserved to a good accuracy or due to some accidental scales mismatch. A celebrated example is our own proton which does make $\sim{ 20\% }$ of all the cold matter. Its longevity ($\tau(proton)\ge {10^{38}}$ seconds), is naturally explained GUT - grand unified theory. It reflects the large ratio $M(Gut)/{m(proton)}/sim {10^{16}}$ in the lifetime predicted; $\tau(p)\sim 1/m(p) ( M(Gut)/{m(p)} )^4$.\\

There are however many cases of well motivated BSM scenarios with no new long lived particles. A well known case involves a  heavy right handed Majorana neutrino which is an $SU(2)_{L}\times U(1)$ singlet. It naturally explains the lightness of the ordinary left handed neutrinos via the see saw mechanism where : $ m_{\nu}(light) \approx {m_D^2/{M_R}}$. With $m_{\nu}(light) \approx{ 0,003-1 eV}$ and a Dirac mass $m_D\approx{ few MeV- 100 GeV}$, the right handed neutrino mass $M_R$ can range from $10^6 GeV $ to $10^{14} GeV$. A stable $\nu_R$ would definitely exceed all the S matrix inspired bounds. However the Yukawa coupling of the Higgs to $\nu_{R}$ and $\nu_{L}$, $Y \bar{\nu}_{R} H \nu (light)$ which through the vev $\langle H \rangle =250 GeV$ supplies Dirac mass $m_D=Y \langle H \rangle$, automatically generates a fast $\nu_{R}\rightarrow{H+\nu_{left}}$ decay.

Also no heavy particle stable arises at the P.Q. ( Peccei Qhinn ) scale of $\approx 10^{10} GeV$,  where the broken $PQ$ symmetry generates the putative axion or in left right symmetric schemes where a new right handed scale  $M(W_R) \ge  \ge M(W_{L})$ nor in a many composite dynamical models.

The option that we discussed at length is to drastically enhance the annihilation rate by endowing the X particles with new interactions. While new ``dark photon" generated forces only slightly relax the bounds non abelian confining interactions drastically improve the situation. At the confining phase transitions the X particles form $\bar{X}-X$ pairs connected with analogs of the chromoelectrix fluxon/string in QCD and a late efficient annihilation stage then results. The simple ``SIMP" type approach  of having the X carry ordinary color has to face he difficulty of forming super heavy isotopes.

The $\sim{0.3\ GeV}$  mass scale of QCD and the $\approx{0.3 TeV}$  scale of E.W symmetry breaking are well known. On the high end new physics regularizing gravity is expected at or before the Planck mass and  GUT theories unifying the $SU(3)XSU(2)XU(1)$ gauge groups and the quarks and leptons of the SM at a high scale of $\approx{10^{16} GeV}$ are very attractive.  A fundamental question is wether between these two ends we have a huge gap or ``desert" or many intermediate scales associated with new physics in between. Without a breakthrough in accelerator technology we will not be able to directly explore this experimentally. The LHC  discovery of  a ``light" 125 GeV  Higgs with no indication of an associated supper-symmetry or new dynamics makes the desert scenario (at least for the next decades of energy) more credible. What our above discussion suggests is that while an extended desert is \textit{not} mandatory it may be quite natural. 

\begin{acknowledgements}
I would like to acknowledge N. Arkani-Hamed, K. Blum, Z. Chacko, P. Saraswat, G. Stermann and M. Zaldarriaga for helpful discussions. 
\end{acknowledgements}

\newpage
\begin{figure}
\begin{center}
\includegraphics[scale=0.7]{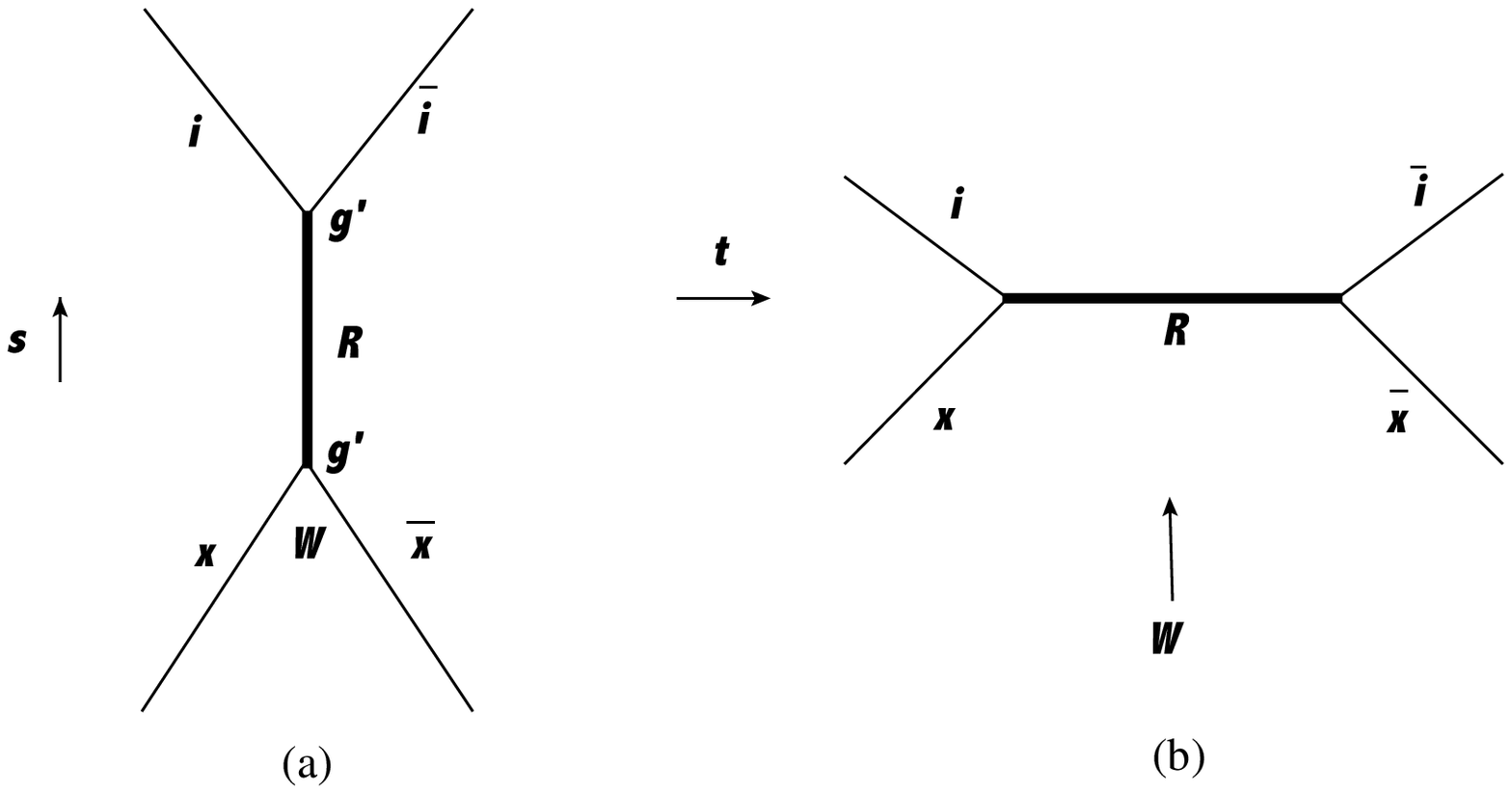}
\caption{(a) A resonance of a bound state in the s-channel (b) Heavy particle exchange in the t-channel}
\label{Diagram1}
\end{center}
\end{figure}

\begin{figure}
\begin{center}
\includegraphics[scale=0.7]{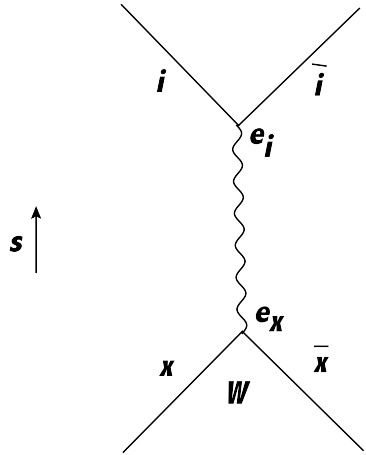}
\caption{An ``elementary" vector boson exchange in the s-channel}
\label{Diagram2}
\end{center}
\end{figure}

\begin{figure}
\begin{center}
\includegraphics[scale=0.7]{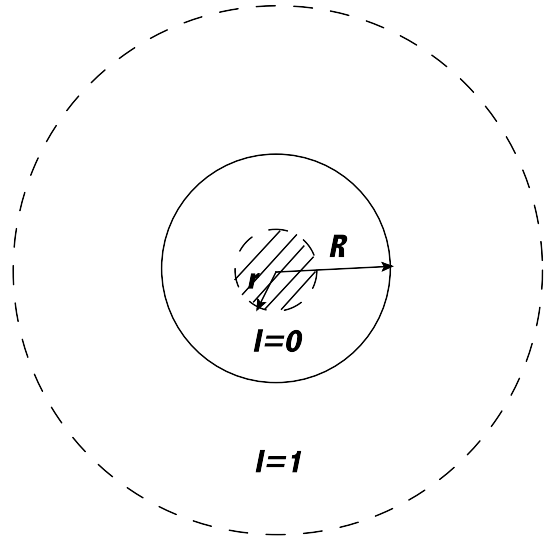}
\caption{The range for scattering and the smaller range for annihilation}
\label{Diagram4}
\end{center}
\end{figure}

\begin{figure}
\begin{center}
\includegraphics[scale=0.7]{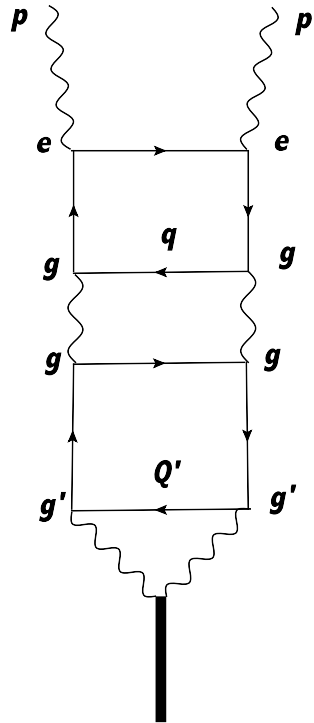}
\caption{The highly suppressed decay of the $g'g'$ glueball into two photons via double box diagram}
\label{Diagram3}
\end{center}
\end{figure}

\end{document}